# Visualizing mitochondrial $F_oF_1$-ATP synthase as the target of the immunomodulatory drug Bz-423

Ilka Starke[1,2], Gary D. Glick[3], Michael Börsch[1,4]

[1] Single-Molecule Microscopy Group, Jena University Hospital, Friedrich Schiller University, Nonnenplan 2 - 4, 07743 Jena, Germany

[2] Institute for Physical Chemistry, Albert Ludwig University Freiburg, Albertstrasse 23a, 79104 Freiburg, Germany

[3] Department of Chemistry, University of Michigan, 930 N University Ave, Ann Arbor, MI 48109-1055, USA

[4] Abbe Center of Photonics (ACP), Jena, Germany

**Abstract**

Targeting the mitochondrial enzyme $F_oF_1$-ATP synthase and modulating its catalytic activities with small molecules is a promising new approach for treatment of autoimmune diseases. The immuno­modulatory compound Bz-423 is such a drug that binds to subunit OSCP of the mitochondrial $F_oF_1$-ATP synthase and induces apoptosis *via* increased reactive oxygen production in coupled, actively respiring mitochondria. Here we review the experimental progress to reveal the binding of Bz-423 to the mitochondrial target and discuss how subunit rotation of $F_oF_1$-ATP synthase is affected by Bz-423. Briefly, we report how Förster resonance energy transfer (FRET) can be employed to colocalize the enzyme and the fluorescently tagged Bz-423 within the mitochondria of living cells with nanometer resolution.

## 1. Introduction

Cellular processes as metabolism and transport are powered by the universal chemical "energy currency" that is the molecule adenosine triphosphate (ATP). Therefore, millimolar ATP concentrations inside the cell have to be produced and maintained through sequential catalytic reactions by the glycolysis pathway or more efficiently by the $F_oF_1$-ATP synthase as part of the oxidative phosphorylation (OXPHOS) pathway. $F_oF_1$-ATP synthases are working in the plasma membrane of bacteria or in small organelles inside of eukaryotes, i.e. the thylakoid membrane in chloroplasts or the inner mitochondrial membrane. In case of mitochondrial $F_oF_1$-ATP synthase a proton motive force (PMF) comprising a concentration difference of protons across the membrane ($\Delta$pH) plus an electric membrane potential ($\Delta\psi$) is required for ATP synthesis. The PMF is generated by sequential redox processes and associated proton pumping of the enzyme complexes I to IV of the respiratory chain across the inner mitochondrial membrane.

If we consider autoimmune diseases, for example systemic lupus erythematodes, being caused by hyperactivity of pathogenic T cells of the immune system, then controlling their cellular ATP concentration with drugs and reducing their activity could become a promising approach for clinical treatment. Modulating T cell activity temperately could circumvent a complete shut-down of the normal immune function. Therefore, one option would be controlling the PMF by targeting any of the enzyme complexes I to IV of the respiratory chain with inhibitors. *Vice versa*, controlling the efficiency of converting PMF to ATP synthesis by $F_oF_1$-ATP synthase would be a possible approach. This latter process is called uncoupling.

More than a decade ago, a 1,4-benzodiazepine, Bz-423 (Figure 1 A), has been found to target lymphoid cells in a murine model of lupus erythematodes[1]. Bz-423 specifically induced apoptosis of pathogenic lymphocytes and attenuated disease progression. As a result, the treated mice showed a prolonged survival at the therapeutic dosage without adverse toxicity and with maintained immune function[2]. The mechanism of Bz-423 action was revealed and subsequently the molecular target was identified – the mitochondrial $F_oF_1$-ATP synthase[3]. Here we focus on the discovery of the drug binding site and discuss a recent microscopy approach using Förster resonance energy transfer (FRET) that has directly demonstrated the binding of a fluorescent Bz-423 derivative to the mitochondrial enzyme in living cells[4].

## 2. Bz-423 binds to OSCP of mitochondrial $F_oF_1$-ATP synthase

Initially, Bz-423 was identified as a potential drug candidate from a library of 1,4-benzodiazepines generated by diversity-oriented chemical synthesis[1]. Phenotype screening of Ramos B cells revealed that Bz-423 caused cell shrinkage, nuclear condensation, cytoplasmic vacuolization, membrane blebbing and DNA fragmentation. Other 1,4-benzodiazepine derivatives were found to selectively target T-cells[5-8]. Bz-423 did not strongly bind to the peripheral benzodiazepine receptor. Cytotoxicity of Bz-423 was related to rapidly generated superoxide ($O_2^-$) in mitochondria. Superoxide is one of the reactive oxygen species (ROS) that can chemically damage cellular macromolecules at higher concentrations. However, Bz-423 superoxide signaling for induced apoptosis was proven as the underlying mechanisms[9, 10].

In the presence of 1 mM sodium azide, the proapoptotic $O_2^-$ generation by Bz-423 was abolished[1]. Because sodium azide is an inhibitor of cytochrome c oxidase, i.e. complex IV of the mitochondrial respiratory chain, Bz-423 was proposed to bind to a mitochondrial OXPHOS protein. Binding of Bz-423 did not alter or collapse the electric potential $\Delta\psi$ across the inner mitochondrial membrane. The superoxide response by Bz-423 required active mitochondria in state 3 respiration, but not mitochondria in state 4 with minimal respiration and in the absence of ADP. Comparing the superoxide generation mechanism induced by oligomycin that induces a state-3-to-4 transition of mitochondrial respiration[11] lead to the hypothesis that Bz-423 might cause a state-3-to-4 transition as well and might bind to $F_oF_1$-ATP synthase.

The validation of $F_oF_1$-ATP synthase as the mitochondrial target of Bz-423 was achieved by phage display screening[3]. Briefly, the oligomycin sensitivity conferring protein (OSCP) being a subunit of $F_oF_1$-ATP synthase was determined. Subsequently the binding site of Bz-423 was located by NMR spectroscopy using the isolated subunit OSCP in solution[12]. Figure 1 A shows the model of bovine heart $F_oF_1$-ATP synthase with the highlighted subunit OSCP in green and the amino acid residues involved in binding of Bz-423 as red dots. Accordingly Bz-423 binds to the top of the membrane enzyme at the interface of OSCP with one pair of $\alpha,\beta$ subunits (blue in Fig. 1 A) of the $F_1$ part. Binding of water-soluble Bz-423 analogues to OSCP in a chemical shift perturbation NMR measurement revealed specific residues 51, 55, 65, 66, 75, 77 and 92 that might form a hydrophobic pocket to accommodate the drug. Furthermore, Bz-423 binding resulted in a conformational rearrangement of helices in OSCP and might alter the interaction of OSCP with $F_1$ in an allosteric manner[12].

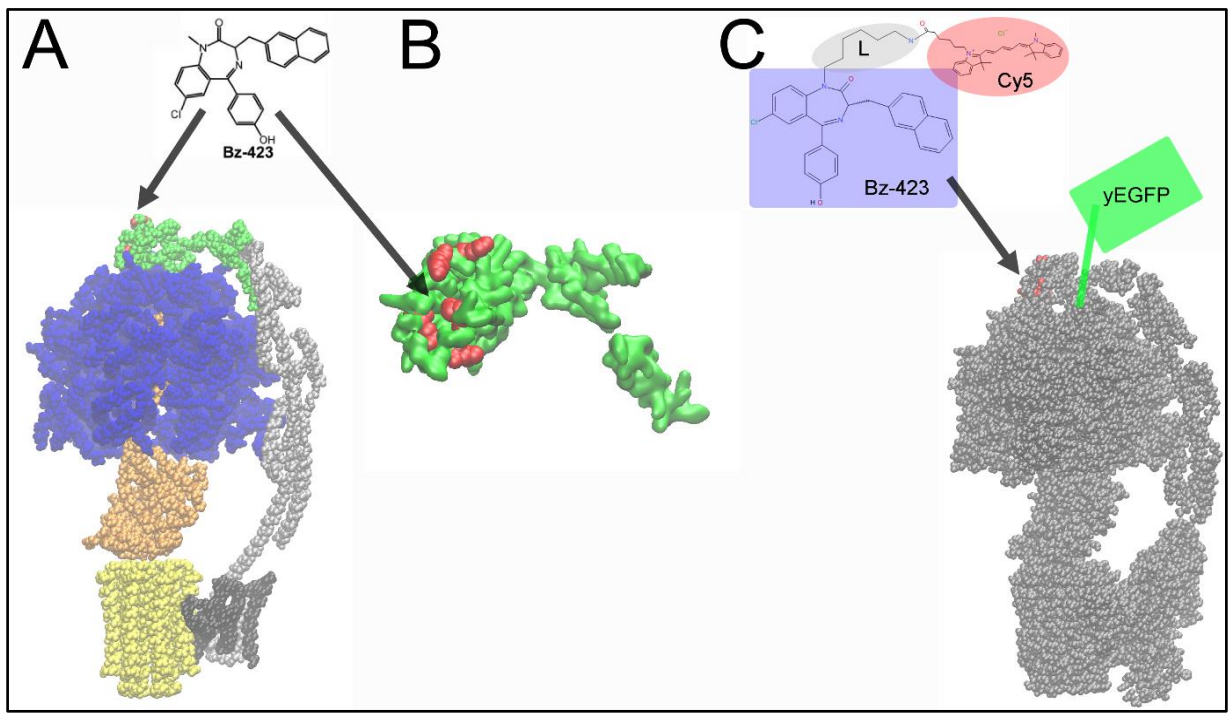

**Figure 1: A**, model of the mitochondrial $F_oF_1$-ATP synthase from bovine heart with binding site for Bz-423 (red dots) on subunit OSCP (in green, on the top; from cryoEM data, PDB 5ARA [13]). Subunits $\alpha$ and $\beta$ are shown in blue and the rotary subunits $\gamma$, $\delta$ and $\varepsilon$ in orange of the $F_1$ part. The peripheral stator consists of subunits *b*, *d*, *$F_6$* and *A6L* (in silver). The proton-translocating subunit *a* is depicted in black, and the rotor ring of eight *c* subunits is shown in yellow. **B,** structure of OSCP from bovine heart (from x-ray crystallographic data, PDB 2WSS [14]) with highlighted residues 51, 55, 65, 66, 75, 77 and 92 shown to be involved in Bz-423 binding. **C,** Bz-423 derivative with the 1,4-benzodiazepine moiety highlighted by the bluish box, a flexible hexyl linker L (grey ellipse) and the FRET acceptor fluorophore Cy5 (red ellipse). Bz-423-Cy5 is expected to bind to OSCP of the yeast mitochondrial FoF1 (red dots on the top of the monomer of the yeast enzyme, cryoEM data, PDB 6B8H [15]). The FRET donor yeast-enhanced green fluorescent protein (yEGFP) is fused to the extended C-terminus of the $\gamma$-subunit (symbolized by the green box).

### 3. Bz-423 requires OSCP to modulate $F_oF_1$-ATP synthase activity *in vitro* and in cells

Binding of Bz-423 to OSCP in the intact $F_oF_1$-ATP synthase inhibited both synthesis and hydrolysis of ATP in isolated sub-mitochondrial particles (SMP) *in vitro*[3, 16]. Both maximal turnover $V_{max}$ and $K_M$ were changed, in contrast to the inhibitor oligomycin which reduced $V_{max}$ only. ATP hydrolysis by the soluble mitochondrial $F_1$ part (comprising the blue and orange colored subunits the Fig. 1 A) was reduced but only when $F_1$ was assembled with OSCP. The $IC_{50}$ for Bz-423 was about 5 μM. In perfused HEK cells ATP synthesis rates of mitochondria were reduced, with $IC_{50}$ of less than 5 μM for Bz-423. Engineered HEK cells with a specifically reduced content of OSCP by siRNA showed alleviated apoptosis by Bz-423, and the residual amount of cellular OSCP correlated well with the percentage of apoptotic cells[3].

$F_oF_1$-ATP synthase accomplishes ATP synthesis by mechanochemical energy conversion with two rotary subunit motors. The PMF drives the $F_o$ motor when protons enter the half-channel of the membrane-embedded *a* subunit (black in Fig. 1 A) from the intermembrane space, i.e. from below. Binding to a specific residue on one *c* subunit (yellow in Fig. 1A) compensates electrostatic forces and allows the ring of *c* subunits to rotate one step forward. Rotation of the *c*-ring moves the elastically-coupled attached $F_1$ motor (orange in Fig. 1 A comprising subunits $\gamma$, $\delta$, and $\varepsilon$). The $F_1$ motor rotates in three 120° steps at high PMF and stops at each of the three catalytic sites where ATP is synthesized in $F_1$. These distinct step sizes of the rotary $F_o$ and $F_1$ motors during ATP synthesis have been measured *in vitro* in single-molecule experiments using bacterial $F_oF_1$-ATP synthases[17-19]. $F_oF_1$-ATP synthase can also run in reverse by hydrolyzing ATP. ATP hydrolysis has been used to investigate the $F_1$ motor in great detail since 20 years[20] and revealed torque, elastic domains[21, 22], substeps[23, 24], breaks and other motor properties[25-27].

Internal subunit rotation with high torque requires a mechanically stable stator counterpart of the enzyme. The stator of the mitochondrial $F_oF_1$-ATP synthase comprises the six $F_1$ subunits $\alpha_3\beta_3$ (blue in Fig. 1 A), subunits *b*, *d*, *F6*, *A6L* (silver in Fig. 1 A), the *a* subunit (black in Fig. 1 A) and OSCP (green in Fig. 1 A). Binding of Bz-423 to the interface of OSCP with $\alpha_3\beta_3$ might weaken the stator assembly and might cause transient uncoupling of the $F_1$ and $F_o$ motors. Alternatively, Bz-423 might influence the subtle conformational changes of OSCP bound to the top part of the catalytic $\alpha_3\beta_3$ subunits and thereby provokes reduced rates of ATP synthesis and hydrolysis. Quantitative enzymatic analysis revealed that Bz-423 is an uncompetitive inhibitor of mitochondrial $F_oF_1$-ATP synthase decreasing $V_{max}$ for ATP synthesis to 50% in the presence of ~10 μM Bz-423[16]. Inhibition by μM amounts of Bz-423 corresponded to fast off-rates < 0.3 $s^{-1}$ of the drug from $F_oF_1$-ATP synthase and a 90% recovery of ATP synthesis rates after 10 seconds.

### 4. Imaging the drug and its molecular target in mitochondria

Localizing a drug at a specific target in life cells can be achieved using fluorescence microscopy with high spatial and temporal resolution as well as single molecule sensitivity. A variety of functional Bz-423 derivatives was synthesized with a flexible linker to rhodamine- or cyanine fluorophores, for example Bz-423-Cy5 (Fig. 1 C). Because μM concentrations of Bz-423 are required to bind to OSCP and to induce apoptosis, direct imaging of the fluorescent drug bound to $F_oF_1$-ATP synthase in the inner mitochondrial membrane is not possible due to a high fluorescent background of unbound Bz-423 throughout the cell. Fast off-rates of Bz-423 and its fluorescent derivatives prevent extensive washing of the cells which is needed to obtain a good imaging contrast. Therefore, confocal microscopy with about 250 nm resolution or superresolution microscopies like structured illumination SIM

[28](resolution limit of about 100 nm) or stimulated emission depletion STED [29](resolution limit of about 20 nm) cannot be used to identify bound Bz-423 on OSCP.

Instead, Förster resonance energy transfer[30, 31] (FRET) as a distance measurement approach in the 2 to 9 nanometer range is applicable. The dipole-dipole interaction of FRET between to nearby fluorophores causes a relative loss of fluorescence intensity of the FRET donor (excited by the laser) and an increase of fluorescence intensity of the FRET acceptor. Thus FRET can be used to relate the spatial position of the fluorescently labelled drug to its cellular target that is tagged with a different fluorophore. Mitochondrial $F_oF_1$-ATP synthase can be tagged with fluorescent proteins, for example at OSCP[32] or at the rotary $\gamma$ subunit without affecting function[33-35]. The benefit of using a genetic fusion to the enzyme is a negligible fluorescent background from other parts of the cell than the cristae of the inner mitochondrial membrane.

5. **Revealing Bz-423-Cy5 binding to $F_oF_1$-ATP synthase by FRET acceptor photobleaching**

To detect binding of fluorescent Bz-423 to mitochondrial $F_oF_1$-ATP synthase in living *S. cerevisiae* cells, we used a $F_oF_1$-ATP synthase mutant designed by J. Petersen and P. Gräber (Albert Ludwig University Freiburg, to be published). A fusion of the yeast-enhanced green fluorescent protein (yEGFP) linked to the C-terminus of the $\gamma$ subunit (Fig. 1 C) provided the donor for FRET imaging (Fig. 2). The fully functional mutant was checked for ATP synthesis, and catalytic rates were also determined *in vitro* after protein purification and reconstitution to liposomes. Staining mitochondria was achieved by incubating the *S. cerevisiae* cells with 4 µm Bz-423-Cy5 (see structure in Fig. 1 C) in the presence of 2% EtOH for 8 h at 28°C. After washing, the bluish cells (Fig. 2 E) were imaged immediately on the microscope at 22°C[4].

Widefield fluorescence microscopy of the stained yeast showed spherical cells with bright spots suggesting fluorescent $F_oF_1$-ATP synthase in mitochondria (Fig. 2 A). SIM imaging of unstained yeast confirmed that only mitochondria were fluorescent[4]. Repeated imaging of the same field of view indicated comparable pixel intensities in these cells revealing only minor photobleaching of yEGFP. However, subsequent excitation of the cells with 640 nm at high power photobleached the FRET acceptor Cy5 dye on Bz-423, and the loss of FRET acceptor resulted in an increase of the FRET donor intensity when imaged again with 488 nm (Fig. 2 B). Stepwise photobleaching of Cy5 correlated with an stepwise increase of FRET donor fluorescence (Fig. 2 C, D). Analysis of the intensity increase in individual cells (see histograms of the single cell highlighted by the red circle in Fig. 2 A-D) due to Cy5 photobleaching unequivocally corroborated the binding of Bz-423-Cy5 to a position on $F_oF_1$-ATP synthase only few nm away from the yEGFP chromophore at the extended C-terminus of the $\gamma$ subunit. Most likely, Bz-423-Cy5 was bound to OSCP.

FRET acceptor photobleaching to confirm the molecular target of a drug in living cells is a fluorescence microscopy approach that requires fast imaging capability but not necessarily high photon counts rates per pixel to achieve accurate colocalization. Despite weak (µM) binding affinities of Bz-423-Cy5 and fast exchange of the immunomodulator on OSCP, a significant fraction of bound Bz-423-Cy5 was revealed by FRET. Slow transport of Bz-423 across the membranes into the matrix of mitochondria as the final destination was indicated by long incubation times needed for staining the yeast cells (Fig. 2 E). Accordingly, a significant fraction of Bz-423-Cy5 still remained in the cytosol and outside of the mitochondria during FRET acceptor photobleaching, and did not contribute to apoptotic action. However, slow exchange of photobleached Bz-423-Cy5 into and out of the matrix compartment facilitated the FRET detection.

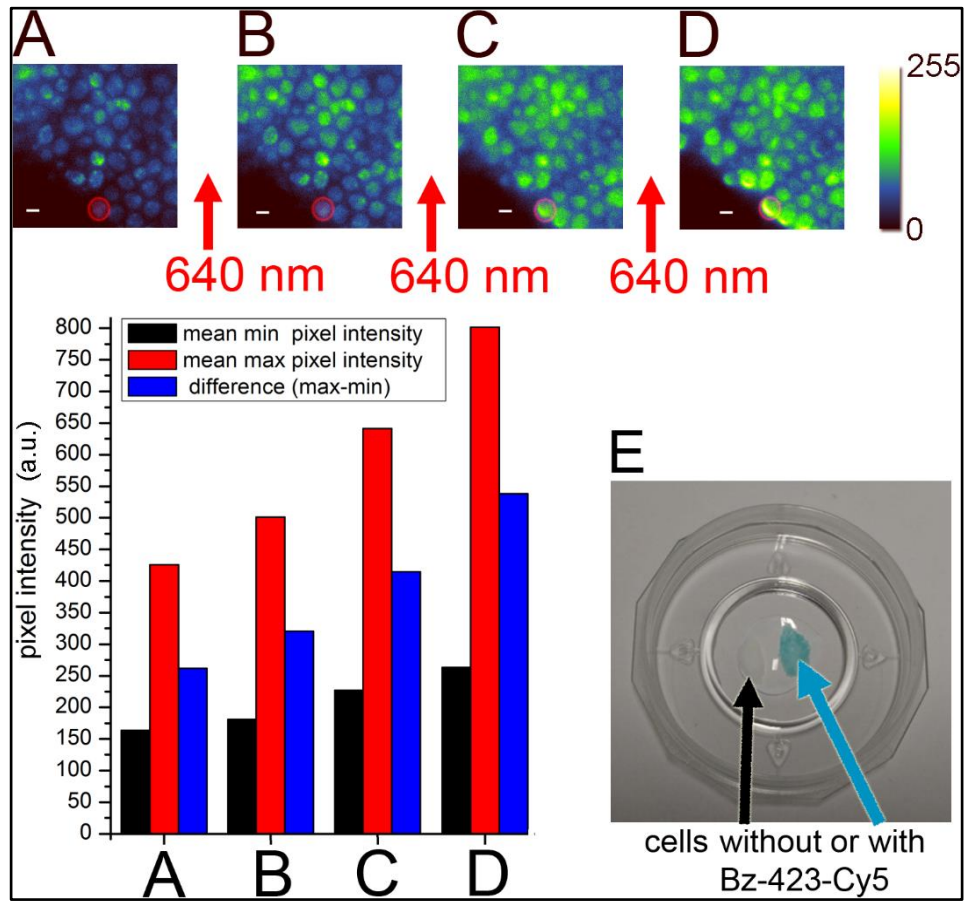

**Figure 2:** FRET acceptor photobleaching of Bz-423-Cy5 bound to mitochondrial $F_oF_1$-ATP synthase with yEGFP fusion to the $\gamma$ subunit in living *S. cerevisiae*[4]. EMCCD-based widefield fluorescence microscopy used laser excitation of yEGFP with 488 nm and fluorescence detection of yEGFP between 500 to 550 nm. **A – D**, sequential imaging of yEGFP-tagged $F_oF_1$-ATP synthases with recalculated, false-colored intensities in 8 bit (0 – 255). Between each image, a 30-s high-power laser pulse with 640 nm was applied to partially photobleach the Cy5 chromophore. The individual pixel intensities of the highlighted cell (red circle as the region-of-interest ROI) are plotted in the histogram in the lower panel as recorded by the EMCCD camera. **E,** comparison of *S. cerevisiae* cells without Bz-423 on the left and stained with 4 µM Bz-423-Cy5 on the right (modified from [4] with permission).

### 6. Future developments

Following the first demonstration that the 1,4-benzodiazepine Bz-423 induced apoptosis in living B cells by stimulating superoxid production of the OXPHOS complexes[1] the identification of its molecular target in mitochondria was accomplished by biochemical methods. Subunit OSCP of $F_oF_1$-ATP synthase being the destination of the drug was unraveled by human cDNA T7 phage display screening. Genetic removal of OSCP in mutant $F_oF_1$-ATP synthases proved that apoptosis required binding of Bz-423 to this subunit. Using purified soluble OSCP, the amino acids involved in the binding site were discovered by NMR spectroscopy[12].

A detailed view on the Bz-423 binding site at the interface of OSCP with the N-termini of $\alpha$ and $\beta$ subunits of $F_1$ is permitted by the recent high-resolution CryoEM structures of the complete mitochondrial enzymes from bovine heart and from yeast *S. cerevisiae*[13, 15]. Both OSCP and the N-termini of $\alpha$ and $\beta$ subunits change their conformations during catalysis, and Bz-423 binding might interfere with these changes and might retard the turnover. Subsequently the mitochondrial PMF builds up, mitochondria switch from respirational state 3 to 4, superoxide is produced, and $Ca^{2+}$ signaling and apoptosis are initiated. Bz-423 induced the opening of the mitochondrial permeability transition pore and thus decreased the $Ca^{2+}$ retention capability[36-38].

The weak binding affinities of Bz-423 and its fluorescent derivatives like Bz-423-Cy5 prevented direct fluorescence microscopy approaches in living cells. In initial confocal microscopy experiments we noticed a broad spatial distribution of Bz-423-Cy5 in *S. cerevisiae* cells but not a specific staining of the mitochondria. Therefore we evaluated the use of sub-mitochondrial particles with fluorescently tagged $F_oF_1$-ATP synthase for FRET *in vitro* but failed to detect sensitized FRET acceptor emission due to the high fluorescent background of unbound Bz-423-Cy5. The solution for a FRET-based direct detection of Bz-423 binding to OSCP was FRET acceptor photobleaching in mitochondria of living cells. Here the pool of unbound Bz-423-Cy5 is limited, and photobleaching Cy5 with 640 nm at high power is possible without destroying the FRET donor fluorophore yEGFP on $F_oF_1$-ATP synthase. Now, similar FRET experiment are under way with human HEK cells and by using the brightest and more photostable green fluorescent protein mNeonGreen[39] fused to the C-terminus of the γ subunit of $F_oF_1$-ATP synthase[34].

To improve specificity of the cellular distribution of the hydrophobic Bz-423 derivatives and to accelerate an uptake into mitochondria, attaching cationic dyes could be used. As shown by H. W. Zimmermann and coworkers[40-43], almost all fluorophores being lipophilic cations transfer quickly to the inner mitochondrial membrane. There, they can bind to proteins. One identified target was cytochrome C oxidase. This finding could be used to specifically induce photodamage by singlet oxygen as a reactive oxygen species[44-46] for Photodynamic Therapy. Beside photoaffinity labeling approaches, time-resolved FRET was applied to reveal that cytochrome C oxidase was a binding site of these lipophilic cationic photosensitizers acting as FRET donors[46, 47]. Similarly, confocal imaging FRET donor lifetimes of mitochondrial $F_oF_1$-ATP synthases in the presence and absence of FRET acceptor-tagged Bz-423 derivatives could be employed to provide direct optical evidence for Bz-423 binding to $F_oF_1$.

FRET-based direct evidence for Bz-423 acting at the mitochondrial $F_oF_1$-ATP synthase needs to be complemented by detailed analysis of cristae morphology changes. In mitochondria of living cells, this can be accomplished by superresolution microscopy approaches, for example structured illumination microscopy[28] (SIM) or stimulated emission depletion[48] (STED) microscopy. Especially, a rearrangement of row assembly of dimeric $F_oF_1$-ATP synthases at the rim of the cristae could indicate the beginning and early events of apoptosis. Finally an unequivocal mechanistic demonstration of how Bz-423 affects the catalytic activity and the rotary motors of mitochondrial $F_oF_1$-ATP synthase is awaited. Single-molecule FRET between different fluorophores attached to rotor and stator of the mitochondrial enzyme has already been developed. Accordingly, the inhibition mechanism of Bz-423 could be unraveled by single-molecule rotation experiments, as shown previously for the inhibitor Aurovertin[49] or of phytopolyphenols[50].

**Acknowledgements**

We gratefully acknowledge the collaboration with Mark Prescott (Monash University), Jan Petersen and Peter Gräber (Albert Ludwig University Freiburg) who designed and provided γ subunit mutants of the mitochondrial enzyme of *Saccharomyces cerevisiae*. Kathryn M. Johnson synthesized the Cy5-tagged Bz-423 (University of Michigan). This work was supported by NIH grant AI-47450 (to G.D.G.) and in part by DFG grant BO 1891/15-1 (to M.B.). The N-SIM / N-STORM superresolution microscope was funded by the State of Thuringa (grant FKZ 12026-515 to M.B.).